\documentclass[onecolumn,showpacs,preprintnumbers,amsmath,amssymb]{revtex4}
\usepackage{graphicx}
\begin{document}
\title{Renewal aging and linear response}
\author{Paolo  Allegrini$^{1}$}
\author{Gianluca Ascolani$^{1}$}
\author{Mauro Bologna$^2$}
\author{Paolo Grigolini$^{1,3,4}$}

\affiliation{$^1$Dipartimento di Fisica ``E.Fermi'' - Universit\`{a} di Pisa, Largo
  Pontecorvo, 3 56127 Pisa, Italy}
  \affiliation{$^2$ Departamento de F\'{i}sica, Universidad de Tarapac\'{a}, Campus Vel\'{a}squez, Vel\'{a}squez 1775, Casilla 7-D, Arica, Chile}
\affiliation{$^3$Center for Nonlinear Science, University of North Texas, P.O. Box 311427, Denton, Texas 76203-1427, USA}
\affiliation{$^4$Istituto dei Processi Chimico Fisici del CNR, Area della Ricerca di Pisa, Via G. Moruzzi, 56124, Pisa, Italy}

\date{\today}
\begin{abstract}
We study the linear response to an external perturbation of a renewal process, in an aging condition that, with no perturbation, would yield super-diffusion.  We use the phenomenological approach to the linear response adopted in earlier work of other groups,
and we find that aging may have the effect of annihilating any sign of coherent response to harmonic perturbation. We also  derive the linear response using dynamic arguments and we find a coherent response, although  with an intensity dying out very slowly. In the case of a step-like perturbation the dynamic arguments yield in the long-time limit a steady signal whose intensity may be significantly smaller than the phenomenological approach prediction.
\end{abstract}
\pacs{05.40.Fb,02.50.-r,82.20.Uv}

\maketitle
\section{introduction}\label{introduction}
The problem of linear response of a dynamic system to an external perturbation \cite{linearesponse} is one topic of general interest in the field of statistical thermodynamics. Many authors \cite{manyauthors} have studied the linear response emergence from microscopic chaos, and thus from a condition that is expected to generate ordinary statistical physics \cite{ordinary}.
In the last few years an increasing number of investigators have been addressing the interesting issue of the response to perturbations of dynamical systems departing from the condition of ordinary statistical thermodynamics \cite{labellaelena,einstein}. A great attention has been devoted to the case of sub-diffusion that makes it possible to fulfill the Einstein relation \cite{einstein} even if the dynamic system generates anomalous diffusion \cite{subdiffusion}.

It is important to point out that the same dynamic generator of intermittence  yields either super- or sub-diffusion according to whether the long sojourn times refer to the velocity or to the position state, respectively. Let us consider a regular one-dimensional lattice, with the fixed distance $a$ between the nearest neighbors, and a particle jumping from one to another site of this lattice. Let us imagine that the particle sojourns for a long time $\tau$ in one site and that at the end of this sojourn makes a jump from the position $na$, with $n$ being an integer number ranging from $-\infty$ to $+\infty$,  either to the position $(n+1)a$ or to the position $(n-1)a$, according to the coin-tossing prescription. This condition can be realized with a dynamic generator of events, occurring at times $t_{0} = 0$, $t_{1},t_{2},\cdots, t_{i},\cdots $, with the waiting times $\tau_{i} =t_{i+1}-t_{i}$ corresponding to a given distribution $\psi(\tau)$. Notice that a single jump involves the occurrence of two events. One event is the drawing of a number $\tau$ from the distribution $\psi(\tau)$, and the other event is the coin tossing. It is worthwhile to stress that in this article we refer to time $t$ as an integer number. The adoption of the continuous time picture is a good approximation made possible by the choice of conditions that involve $t \gg 1$. 

With the same dynamic model and the same coin-tossing prescription, we can generate a different random walk process, producing super-diffusion. Here the random walker moves with a velocity of fixed modulus $W$, whose sign is determined by two events occurring at the times $t_{i}$. As in the earlier example, the two events are the drawing of the time $\tau_{i}$ from the distribution $\psi(\tau)$ and the coin tossing that fixes the velocity sign of the walker between time $t_{i}$ and time $t_{i+1}$. This corresponds to defining two states, the state $|1>$, with velocity $W$, and the state $|2>$, with velocity $-W$. We refer to the time interval $\tau_{i} =t_{i+1}-t_{i}$ as \emph{waiting time}. Of course, the walker can repeatedly jump to the right (left), in the first case, or maintain the same positive (negative) velocity sign, in the second case. The two processes are different, insofar as in the first case, the larger the sojourn time the slower the diffusion process, while in the second the opposite condition applies: the larger the sojourn times the faster the diffusion process. In this paper we adopt the velocity rather than the position picture, the reason for this choice being that, as we shall see, it establishes a connection with the interesting phenomenon of non-Poisson stochastic resonance \cite{hanggi}. We shall be referring throughout to the wide time regions between $t_{i}$ and $t_{i+1}$, where no unpredictable events occur, as \emph{quiescent regions}. We shall adopt a dynamic model inspired to turbulence and to the Manneville map \cite{manneville},where the quiescent regions are usually denoted as laminar regions.  This dynamic model \cite{paulthegreat,lastgerardo}
yields the waiting time distribution $\psi(\tau)$ with the inverse power law form:
\begin{equation}
\label{theoretical}
\psi(\tau) = (\mu -1) \frac{T^{\mu -1}}{(\tau+T)^{\mu}},
\end{equation}
with $\mu > 1$ and
\begin{equation}
\label{tocontinuum}
T \gg 1.
\end{equation} 
The condition of Eq. (\ref{tocontinuum}) is essential to adopt, whenever convenient, the earlier mentioned continuum time treatment. The probability that no event occurs until time $t$ is
\begin{equation}
\Psi(t) \equiv \int_{t}^{\infty} dt^{\prime} \psi(t^{\prime}) = \left(\frac{T}{t+T}\right)^{\mu-1}.
\end{equation}
This function is called survival probability.

When $\mu > 2$, the mean value of these waiting times is given by
\begin{equation}
<\tau> = \frac{T}{(\mu -2)}.
\end{equation}
The condition $\mu = 2$, where this mean value diverges is the border between the region $\mu > 2$, compatible with the existence of an infinitely aged condition \cite{infinitelyaged}, and the region $\mu < 2$, which is characterized by perennial aging and ergodicity breakdown \cite{ergodicitybreakdown}. In this paper we focus our attention on the condition of perennial aging $\mu < 2$.

For a detailed discussion of renewal aging we refer the reader to the recent literature on this subject, for instance, Refs. \cite{aging1,aging2}. The proper quantitative treatment of aging requires \emph{preparation}, namely, that the time origin $t=0$ corresponds to an event occurrence. Aging means that the waiting time distribution $\psi(\tau)$ and the corresponding survival probability $\Psi(\tau)$ depend on the time at which the observation process begins. We define with the symbol $\psi(t,t^{\prime})$ the probability of meeting the first event at time $t$ when the observation begins at time $t^{\prime} <t$ and the system is prepared at time $t=0$.  $\Psi(t,t^{\prime})$ denotes the corresponding survival probability. The functions $\psi(t,t^{\prime})$ and $\Psi(t,t^{\prime})$ refer to a condition where
$t=t_{i}$, namely, a time at which an event occurs, and $t^{\prime}$ is a time at which no event occurs and observation begins.  Of course, the Poisson conditions
$\psi(t,t^{\prime}) = \psi(t-t^{\prime})$ and $\Psi(t,t^{\prime}) = \Psi(t-t^{\prime})$ are violated \cite{aging1,aging2} but in the specific case where
$t^{\prime} = t_{i-1}$, namely, when observation begins with an event occurrence. More precisely, the functions $\psi(\tau)$ and $\Psi(\tau)$ refer to the condition $\tau = t_{i} -t_{i-1}$, with $t_{i}$, denoting, according to the earlier definitions, times at which events occur.
To discuss the response of renewal non-Poisson systems to perturbations we  study their behavior in the presence of a time dependent perturbation, a condition that forces us to introduce also the function $\psi(t|t^{\prime})$. This is the conditional probability that at time $t$ an event occurs, given the condition that the earlier event occurs at time $t^{\prime}$. In the limiting case of extremely weak perturbation it is expected that
\begin{equation}
\label{fixedrules}
\psi(t|t^{\prime}) = \psi(t-t^{\prime}).
\end{equation}
In fact, in the unperturbed case, as earlier pointed out, the waiting time distribution is obtained by recording the distance between an event and the next, insofar as any time can be selected as a time origin, if it corresponds to an event occurrence. Renewal aging does not have anything to do with the system being driven by physical rules changing with time, and it is only a consequence
of making observation and preparation at different times. This condition, and the equality of Eq. (\ref{fixedrules}) with it, is violated by external perturbation. Let us note furthermore that in the literature the symbols $\Psi$ and $\psi$ are usually adopted to denote a probability and a probability density, respectively. Due to the adoption of a discrete time representation in this article both symbols denote probabilities, insofar as, for instance, $\psi(\tau) d\tau = \psi(\tau)$, as a consequence of the fact that for the integration time step we have $d\tau = 1$.

 The problem of the response of sub-diffusion to external perturbation in an aging condition has been recently addressed by Sokolov, Blumen and Klafter \cite{sokolov} and by
Barkai and Cheng \cite{barkai}. These authors have adopted the same phenomenological model as Bertin and Bouchaud \cite{bertin}. In this model the external perturbation, which does not affect the sojourn time duration, influences  the choice between either jumping to the right or to the left, at the end of a sojourn: In the unperturbed case there is no bias, this choice being done with a fair coin, and the effect of perturbation is to turn this fair coin into an unfair coin, producing a bias at the moment of the choice between the left and the right nearest-neighbor site.
The phenomenological method will be extended to the velocity picture. However, in addition to adopting the velocity picture, and thus the physical condition of super-diffusion rather than sub-diffusion, we plan to go beyond the phenomenological model. For this reason, we shall also study the case when the perturbation-induced bias is generated through the influence that the external perturbation exerts on
the waiting time duration. Although this is in principle a hard problem, we find an analytical solution in the limiting case of weak perturbation: moving along the lines of the linear response method \cite{linearesponse}, we express the response to perturbation in terms of the system's unperturbed statistical properties. In the aging regime of this paper the dynamic model  yields results that are significantly different from those produced by the phenomenological model, and identical to those of Ref. \cite{prl}, where the dynamical model was already used although with no demonstration \cite{prlnote}. It is important to stress that also the adoption of
the phenomenological model allows us to obtain original results, insofar as it deals with the response of super-diffusion, rather than sub-diffusion \cite{sokolov,barkai}, to an external perturbation.

The outline of this paper is as follows. In Section \ref{dynamic1} we illustrate a dynamic model, generating, in the unperturbed case, a non-Poisson intermittent signal. We shall show that an external perturbation has the effect of perturbing the distribution of waiting times in both velocity states, and we shall discuss how to convert this property into a bias.  Section  \ref{stochasticliouville} illustrates a procedure to derive the response to perturbation based on a master equation with a time-dependent rate. We also write, using a discrete time picture, the exact solution of this stochastic master equation, as a time dependent and stochastic signal. We discuss how to evaluate the statistical averages of this signal in Sections \ref{phen} and \ref{dynamic}. In Section \ref{phen} we evaluate this statistical average by making the phenomenological assumption
that the perturbation-induced bias is determined by the tossing of an unfair coin when a collision occurs. In Section \ref{dynamic} we discuss how to convert, in the weak perturbation limit, the perturbed waiting time into a bias.
Section \ref{asym} illustrates the system's response in the time asymptotic limit, in both the phenomenological and dynamic case. In Section \ref{final} we summarize the main results of this paper. The appendixes A and B illustrate the algebra behind the main results of Section \ref{asym} in the phenomenological and the dynamic case, respectively.

\section{Dynamic model}\label{dynamic1}
In this section we discuss how to go beyond the phenomenological approach, by means of a dynamic model that has been used in earlier work (see \cite{paulthegreat,lastgerardo} and references therein) with different motivations: This dynamic model serves the purpose of emphasizing the limits of the Liouville approach to super-diffusion \cite{paulthegreat} as well as the purpose of providing an efficient dynamic algorithm for the non-Poisson renewal condition \cite{lastgerardo}.  Here we use this  model to simulate the effect of an external perturbation on a non-Poisson dynamic process, with the following procedure. Let us consider the equation of motion:
\begin{equation}
\label{modulatedrenewal}
\dot y = \alpha(t) y^{z},
\end{equation}
with $z> 1$, and $\alpha(t)\ll 1$. As pointed out in Section \ref{introduction}, in this article we adopt a discrete time picture, namely,
\begin{equation}
\label{modulatedrenewal2}
y_{t+1} = y_{t} + \alpha(t) (y_t)^z.
\end{equation}
The condition $\alpha(t)\ll 1$ makes it possible for us to convert the discrete into a continuous time picture, insofar as the event occurrence involves $t \gg1$, and so an integer virtually indistinguishable from a continuous time.
Eq. (\ref{modulatedrenewal}) describes a particle moving within the interval $I=(0,1]$: When the particle reaches the border, $y = 1$, it is injected back to new initial conditions within the interval $I$ with the same probability, $p(y) = 1$. The event of the particle reaching the border $y=1$ coincides with the crucial collision event that will be discussed in the next section.

It is important to point out that we have in mind as an example of non-Poisson two-state system the blinking quantum dots \cite{intermittentfluorescence}. These physical systems are nano-crystals that under the influence of a radiation field produce intermittent fluorescence, which is equivalent to a random transition from a ``light on" state, with the system emitting fluorescent light, to a ``light off" state, with no light emission. The waiting time distribution in these two states is not exponential, and the ``light on" distribution is different from the ``light off" distribution. In this paper we are interested in establishing the bias created by an external perturbation. Thus, we make the assumption that these two distributions, in the absence of perturbation, are identical. In the case when $\alpha(t)$ is time dependent as an effect of an external perturbation, its time evolution may depend on whether the system is in the state ``on" or ``off". Thus, it is convenient to adopt the following notation
\begin{equation}
\label{differentstates}
\alpha_{\pm}(t) = \alpha_{0}[1+ \epsilon F_{\pm}(t)],
\end{equation}
with the subscripts $+$ and $-$ referring to the states $|1>$ and $|2>$, respectively. The parameter $\epsilon$ denotes the perturbation strength.
The collision event triggers a fair coin tossing, which has the effect of fixing the sign of the ensuing state.

It is worthwhile to point out that the dependence of $\alpha(t)$ on the state sign makes it convenient to distinguish the two events. We reserve the term \emph{collision} to the variable $y$ reaching the border $y=1$. We shall explicitly refer to the other event as \emph{coin tossing event}. We shall see that the effect of perturbation ($\epsilon \neq 0$) is to make the collisional  depend on the coin-tossing event. 

In  the case when $\alpha(t) = \alpha_{0}$, namely, when $\alpha_{\pm}(t)$ is time independent, this model coincides with the generator of inverse power law used in Ref. \cite{lastgerardo}. The model used in Ref. \cite{lastgerardo} is renewal, because the initial condition of $y(t)$ after each back injection is selected randomly. Consequently, the ensuing time of sojourn within the interval $I$ does not have any memory of the earlier sojourn times.
When $\alpha$ is time independent, the model yields the waiting time distribution of Eq. (\ref{theoretical}), with
\begin{equation}
T = \frac{\mu -1}{\alpha_{0}},
\end{equation}
and
\begin{equation}
\mu = \frac{z}{z -1}.
\end{equation}
This indicates that perturbing $\alpha$, while keeping fixed the value of $z$, is equivalent to assuming that the external perturbation changes the parameter $T$ only.

Let us use Eq. (\ref{modulatedrenewal}) to determine the time $\tau$ it takes the system, moving from the condition $y(t^{\prime})$ at time $t^{\prime}$, to reach the border. With an easy algebra, we find
\begin{equation}
\label{initialcondition}
y(t^{\prime}) = \left[ 1 + (z-1) \int_{t^{\prime}}^{\tau+t^{\prime}}  \alpha_{\pm}(t^{\prime \prime}) dt^{\prime \prime} \right]^{-1/(z-1)}.
\end{equation}
This expression allows us to determine the conditional probability $\psi_{\pm}(\tau|t^{\prime})$, which is the probability  of meeting a collision at time $t^{\prime} + \tau$ given that two events, the collision and the coin-tossing event, occur at time $t^{\prime}$. With the assumption of uniform back injection, this conditional probability is given by
\begin{equation}
\psi_{\pm}(\tau|t^{\prime}) d \tau = p[y(t^{\prime})]dy(t^{\prime}),
\end{equation}
with $p[y(t^{\prime})] = 1$. Note the adoption of a continuous time picture for these calculations is made legitimate by the condition $1 \ll d\tau \ll T$, which is, in turn, a consequence of $\alpha(t) \ll 1$.  Thus, we obtain
\begin{equation}
\label{roughform}
\psi_{\pm}(\tau|t^{\prime}) = |dy(t^{\prime})/d\tau| = \frac{\alpha_{\pm}(\tau + t^{\prime})}{\left[1 + \frac{1}{\mu -1}\int_{t^{\prime}}^{\tau+t^{\prime}}\alpha_{\pm}(t^{\prime\prime})dt^{\prime \prime}\right]^{\mu}}.
\end{equation}
The waiting time distribution $\psi_{\pm}(\tau|t^{\prime})$ satisfies the normalization condition
\begin{equation}
\label{normalization}
\int_{0}^{\infty} d\tau \psi_{\pm}(\tau|t^{\prime}) = 1.
\end{equation}
 Note that the function $\psi(\tau|t^{\prime})$ can also be written as follows
\begin{equation}
\psi_{\pm}(\tau|t) \equiv \psi_{\pm}(\tau|t^{\prime}),
\end{equation}
with
\begin{equation}
t = t^{\prime} + \tau.
\end{equation}
 We note again that the time when the earlier jump occurred, a coin was tossed to decide whether the system had to be located in the state $|1>$, sign $+$, or $|2>$, sign $-$. Therefore $\psi(\tau|t)$ is the probability that a jump occurs at time $t$ under the condition that the earlier jump occurred at $t-\tau$ and correspondingly a sign choice was made.

Using Eq. (\ref{differentstates}) we write Eq. (\ref{roughform}) in the more convenient form
\begin{equation}
\label{moreconvenientform}
\psi(\tau|t^{\prime})  = \frac{\alpha_{0}\left[1 + \epsilon F_{\pm}(\tau + t^{\prime})\right]}{\left[1 + \frac{\alpha_{0}\tau} {\mu-1} + \frac{1}{\mu -1}\int_{t^{\prime}}^{\tau+t^{\prime}}F_{\pm}(t^{\prime\prime})dt^{\prime \prime}\right]^{\mu}}.
\end{equation}
Let us make some remarks on the physics behind the theoretical prediction of Eq. (\ref{moreconvenientform}).
We note that
\begin{equation}
t^{\prime}  \equiv \tau_{1} + ...+ \tau_{n-1}
\end{equation}
and
\begin{equation}
\tau \equiv \tau_{n},
\end{equation}
with $n-1$ denoting an arbitrary positive integer number.
 The system is prepared at $t = 0$ and Eq. (\ref{moreconvenientform}) expresses the probability of getting a sojourn time of length $\tau$, given the condition that an earlier jump  occurred at a time $t^{\prime}$, as the last of a sequel of $n-1$ events.

In the special case of a harmonic perturbation,
\begin{equation}
F_{\pm}(t) = \pm \cos(\omega t),
\end{equation}
Eq. (\ref{moreconvenientform}) generates
\begin{equation}
\label{exact} \psi_{\pm}(\tau|t^{\prime}) = \frac{(\mu -1)}{T}
\frac{\left\{1 \pm \epsilon \cos[\omega(t^{\prime}+
\tau)]\right\}}{\left\{1 + \tau/T \pm \frac{\epsilon }{\omega
T}\left[\sin(\omega(t^{\prime} + \tau) )-\sin(\omega
t^{\prime})\right]\right\}^{\mu}}.
\end{equation}
Eq. (\ref{exact}) plays a fundamental role for the dynamic approach of this paper. By expanding it up to the first order in the perturbation parameter $\epsilon$, we get
\begin{equation}
\label{notexact2} \psi_{\pm}(\tau|t^{\prime}) \approx \psi(\tau)
\{1 \pm \epsilon \cos[\omega(t^{\prime}+ \tau)]\} \pm
\epsilon\psi_{\mu+1}(\tau,t^{\prime}).
\end{equation}
This first-order approximation maintains the normalization condition, thanks to the second term on the right-hand side of Eq.(\ref{notexact2}), whose explicit expression is of no interest for the discussion of this section. It is enough to say that for $\tau \rightarrow \infty$, this second term is proportional to $1/\tau^{\mu +1}$. This suggests that the significant contribution to the linear response to perturbation depends only on the first term on the right-hand side of Eq. (\ref{notexact2}), thereby leading to
\begin{equation}
\label{notexact} \psi_{\pm}(\tau|t^{\prime}) \approx \psi(\tau)
\{1 \pm \epsilon \cos[\omega(t^{\prime}+ \tau)]\}.
\end{equation}

The approximated expression of Eq. (\ref{notexact}) suggests a way to convert the exact conditional probability of Eq. (\ref{exact}) into the joint probability of event occurrence and sign drawing. To properly discuss this issue, let us denote by $A_{t}$ and $B_{t}$ the coin tossing and collision events, respectively, occurring at time $t$. Using the conventional probabilistic notations we express $\psi_{\pm}(\tau|t^{\prime})$ of Eq. (\ref{exact}) in the following form
\begin{equation}
\label{conditional}
\psi_{\pm}(\tau|t^{\prime})  = P(B_{t} |A_{t^{\prime}}, B_{t^{\prime}}).
\end{equation}
In fact, as earlier pointed out, $\psi_{\pm}(\tau|t^{\prime})$ is the probability that a collision occurs at $t$ ($B_{t}$), given that fact that the earlier collision and the sign choice of the quiescent region occur at $t^{\prime}$ ($A_{t^{\prime}}$ and $B_{t^{\prime}}$). 
The probability that a collision in a state chosen with the coin tossing triggered by the earlier collision ($A_{t^{\prime}})$ occur at time $t$, under the condition that the earlier collision occurs at time $t^{\prime}$, $P(A_{t^{\prime}}, B_{t} |B_{t^{\prime}})$, fulfills the following equation
\begin{equation}
\label{general}
P(A_{t^{\prime}}, B_{t} |B_{t^{\prime}}) = P(B_{t}|A_{t^{\prime}}, B_{t^{\prime}}) P(A_{t^{\prime}}|B_{t^{\prime}}).
\end{equation}
On the other hand, the use of a fair coin yields
\begin{equation}
P(A_{t^{\prime}}|B_{t^{\prime}}) = P(A_{t^{\prime}}) = \frac{1}{2}.
\end{equation}
Thus, using Eq. (\ref{conditional}), we obtain
\begin{equation}
\label{ascloseaspossible}
P(A_{t^{\prime}}, B_{t}|B_{t^{\prime}}) = \frac{1}{2} \psi_{\pm}(\tau|t^{\prime}).
\end{equation}
This is an exact result, and the adoption of the dynamic approach should keep us as close as possible to it.

As mentioned in Section \ref{introduction}, the phenomenological approach to the response of a walker to perturbation rests on assuming that the choice between jumping to the left and jumping to the right is done with a biased coin. In the case discussed in this article, this corresponds to selecting the state, either $|1>$ or $|2>$, at the beginning of a quiescent region. Thus,  with the phenomenological approach $P(B_{t}|A_{t^{\prime}}, B_{t^{\prime}}) = P(B_{t})$ insofar the occurrence of a collision at time $t$ does not have any dependence on the earlier sign choice, which does not affect the quiescent region time duration. Note that the collision occurring at time $t$ is the first one after the collision at time $t^{\prime}$. Thus, due to the renewal character of the process, $P(B_{t})$ is proportional to $\psi(t-t^{\prime})$. On the other hand, given the fact that the phenomenological approach rests on the sign choice with an unfair coin, $P(A_{t^{\prime}}|B_{t^{\prime}})$ is not independent of $B_{t^{\prime}}$. Thus, we replace Eq. (\ref{general}) with
\begin{equation}
\label{phenomenal}
P(A_{t^{\prime}}, B_{t}|B_{t^{\prime}}) = P(B_{t}) P(A_{t^{\prime}}|B_{t^{\prime}}).
\end{equation}
This yields
\begin{equation}
P(A_{t^{\prime}}, B_{t}|B_{t^{\prime}}) =  \psi(\tau) p_{\pm}(t^{\prime}),
\end{equation}
where
\begin{equation}
\label{probabilities2} p_{\pm}(t^{\prime}) = \frac{1 \mp \epsilon
\cos (\omega t^{\prime})}{2},
\end{equation}
which indicates that we identify $P(A_{t^{\prime}}|B_{t^{\prime}})$ with the unfair-coin tossing event. In other words, the sign choice is done at the beginning of 
the quiescent region, tossing an unfair coin. Note the sign inversion that has the effect favoring the state
$|2>$ ($|1>$) in the same way as the collision occurring in the state
$|1>$ ($|2>$) earlier than in the state $|2>$ ($|1>$). In other words, as an
effect of the approximation of setting $\epsilon = 0$ in the
denominator of the term in the right-hand side of Eq.
(\ref{exact}), with the phenomenological approach we assume that the
conditional probability $\psi(\tau|t^{\prime})$ of Eq.
(\ref{exact}) yields the joint probability $\psi(\tau)
p_{\pm}(t^{\prime})$, which is the probability that at time
$t^{\prime}$ we assign the sign $+$ or $-$ to a quiescent region
of time duration $\tau$. The two events, the drawing of time
$\tau$ and the sign selection, are totally independent. As a
consequence, there is no correlation between sign and time
duration, a fact in a striking conflict with the well known fact
that stochastic resonance can establish a strong correlation
between sign and time duration. The reader can consult Ref.
\cite{moss} for a case of strong departure of the perturbed
waiting time distribution from the unperturbed form. 

Let us consider now the case where the biased sign choice is made at the end of the quiescent region, namely, the case 
\begin{equation}
P(A_{t}, B_{t} |B_{t^{\prime}}) = \psi(\tau)  p_{\pm}(t),
\end{equation}
with
\begin{equation}
\label{probabilities} p_{\pm}(t) = \frac{1 \pm \epsilon
\cos (\omega t)}{2}.
\end{equation}
Note that in this case there is no sign inversion. This choice is equivalent to 
\begin{equation}
\label{puttana}
P(A_{t}, B_{t}|B_{t^{\prime}}) = \psi_{\pm}(\tau|t^{\prime})  p_{\pm}(t),
\end{equation}
where $\psi_{\pm}(\tau|t^{\prime})$ is given by Eq. (\ref{notexact}) and
\begin{equation}
\label{puttanina}
p_{\pm}(t) = \frac{1}{2}. 
\end{equation}
Thus the adoption of this choice allows us to maintain
the promise of remaining as close as possible to the exact dynamic prescription of Eq. (\ref{ascloseaspossible}). This can be realized in two distinct but equivalent ways. With the first prescription, we toss a fair coin either at beginning or at the end of the quiescent region, whose time duration, as a perturbation effect, depends on whether the system is in the state $|1>$ or $|2>$. With the second prescription we select the sign of the quiescent region, whose time duration is unperturbed, tossing a \emph{unfair coin} at the \emph{end} rather than at the \emph{beginning} of the quiescent region. With this prescription we depart from the phenomenological approach, where the sign choice is done, tossing an unfair coin, at \emph{beginning} of the quiescent region. 

 The dynamic and the phenomenological approach, which coincide in the Poisson case, lead to different predictions in the non-Poisson case, and the departure of one prediction from the other becomes significant especially when the condition $\mu < 2$ applies. This is a consequence of the fact that the adoption of a biased coin, tossed at the beginning rather than at the end of the quiescent region, does not reflect quite properly the dynamic origin of the response to perturbation.
If the
phenomenological model is thought of as an approximation to the
dynamic model, it is evident that it has to be limited not only to the case
of extremely weak perturbations, as the dynamic approach, but also to cases where the perturbation-induced
waiting time reordering can be totally neglected, so as to fulfill
Eq. (\ref{fixedrules}).


\section{stochastic Liouville equation}\label{stochasticliouville}
Let us consider the ordinary master equation
\begin{equation}
\label{pauli1}
{\bf p}(t+1)- {\bf p} (t) = -\frac{1}{2} {\bf K} {\bf p}(t),
\end{equation}
where
\begin{eqnarray}
\label{cointossing2}
 {\bf K} \equiv \left(\begin{array} {cc}1 & - 1 \\
                         -1 & 1 \end{array}
                         \right)  \end{eqnarray}
and ${\bf p}(t)$ is a vector, whose components $p_{1}(t)$ and $p_{2}(t)$, with $p_{1}(t) + p_{2}(t) = 1$, denote the probability
for the system to be found in the corresponding states $|1>$ and $|2>$, respectively.
We are considering the case of a discrete time $t$. Thus, Eq. (\ref{pauli1}) is nothing but the probabilistic picture corresponding to a fair coin-tossing prescription. We can imagine a particle located, for instance, in the state $|1>$. If the fair coin tossing yields head, the particle remains in this state. If the fair coin tossing yields tail, the particle jumps to the state $|2>$. If the next coin tossing yields tail again, the particle remains in the state $|2>$. Otherwise it jumps back to the state $|1>$. We define the coin-tossing event as \emph{collision}. The time distance between a collision and the next is equal to $1$, in this case. Thus, collisions are not a source of disorder, which, in this case depends only on the coin tossing. 

It is more realistic to assume that the time distance between one collision and the next is not fixed and fluctuates. The leading idea of Continuous Time Random Walk (CTRW) \cite{mw} can be adapted to this simple process by assuming that the time distance between one collision and the next is not constant, and it is characterized by a probability distribution $\psi(\tau)$. In this paper we develop a theory for the non trivial case where the key function $\psi(\tau)$ has a non-exponential form. We shall make explicit calculations for the special case where $\psi(\tau)$ has the form of Eq. (\ref{theoretical}).
Note that the theory developed in this paper makes it possible for us to address also the case $\mu < 2$. In this case, the mean waiting time is infinite, and the system is found to be in a state of perennial aging, insofar as the stationary condition is not admitted in this case \cite{sokolov,barkai,bertin,prl}.

The theory of this paper rests on the connection between the
CTRW and the Stochastic Liouville Equation (SLE) of Kubo \cite{sle1,sle2,kubo}.
Let us devote some attention to this crucial aspect. As a consequence of the time fluctuations between one collision and the next, we write Eq. (\ref{pauli1}) under the following form
\begin{equation}
\label{stochasticpauli}
\frac{d {\bf f}(t) }{dt} = - \frac{r_{0}(t)}{2} {\bf K} {\bf f}(t),
\end{equation}
where ${\bf f}(t)$ is a two-dimensional vector with components $f_{1}(t)$, $f_{2}(t)$ and $f_{1}(t) + f_{2}(t) = 1$. The fluctuating rate $r_{0}(t)$ always vanishes but in the correspondence of a collision, where it gets the value of $1$. The functions $f_{1}(t)$ and $f_{2}(t)$ denote the probability for the particle to be in the corresponding states $|1>$ and $|2>$, as in the earlier case of constant time distance between two consecutive collisions. The reason for this notation change is that, as in the SLE by Kubo \cite{kubo}, $f_{1}(t)$ and $f_{2}(t)$ are  fluctuating probabilities, rather than ordinary probabilities: In the original work of Kubo \cite{kubo} the fluctuating probability $f(x,t)$ is turned into an ordinary probability, obeying the Fokker-Planck equation, by averaging on the stochastic process.

It is important to stress that our approach is based on a discrete time picture, with the integration time step $dt = 1$. Therefore, we define
\begin{equation}
\frac{d {\bf f}(t)}{dt}  \equiv {\bf f}(t+1) - {\bf f}(t).
\end{equation}
Consequently, when a rare collision occurs, Eq. (\ref{stochasticpauli}) produces the same effect as Eq. (\ref{pauli1}). Let us define
\begin{equation}
\label{bigsigma}
\Sigma(t) = f_{1}(t) - f_{2}(t).
\end{equation}
Thus, Eq. (\ref{stochasticpauli}) becomes
\begin{equation}
\label{bigsigma2}
\Sigma(t) = \Sigma(t-1) -r_{0}(t-1) \Sigma(t-1).
\end{equation}
Note that
\begin{equation}
\Sigma(t+1) = 0
\end{equation}
and
\begin{equation}
\Sigma(t+1) = \Sigma(t),
\end{equation}
according to whether at $t$ a collision does or does not occur, respectively. Thus, a collision occurring at $t$ turns $\Sigma(t)$ into the vanishing value, while, with no collision, the value $\Sigma(t)$ is transmitted to the next time step.

The earlier approach is a convenient way to describe the relaxation process in the unperturbed case. To address the problem of the response to external perturbation we are forced to adopt the following generalization:

\begin{eqnarray}
\label{cointossing}
{\bf f}(t+1) - {\bf f}(t)  = -\frac{1}{2}\left[\begin{array} {cc}r_{+}(t) & - r_{-}(t) \\
                         -r_{+}(t)  & r_{-}(t) \end{array}
                         \right]  {\bf f}(t). \end{eqnarray}
                         With $r_{\pm}(t)$ we denote a stochastic process that is almost always vanishing, except at the moment of a collision. If the collision occurs at the same time for both the $|1>$ and $|2>$ state, phenomenological model, the bias is determined by the assumption of Eq. (\ref{probabilities}), which can be interpreted as generating the state dependent rate
                         \begin{equation}
\label{suitablechoice} r_{\pm}(t) = r_{0}(t)[1\pm \epsilon
\cos(\omega t)],
\end{equation}
where $r_{0}(t)$ is a stochastic function of time, which is almost always a vanishing quantity, except in the rare case of a collision, when it gets the value of $1$. The time distance between two non-vanishing values of $r_{0}(t)$ is randomly drawn from the distribution $\psi(\tau)$ of Eq. (\ref{theoretical}).
                        According to the dynamic models,  the collisions in the states $|1>$ and $|2>$ occur at different times, and, consequently, $r_{+}(t)$ and $r_{-}(t)$ are equal to $1$ when the collision occurs.  Let us imagine, for example, that the collision occurs in the state $|1>$. We see that Eq. (\ref{cointossing}) corresponds to shifting  half particles from the state $|1>$ to the state $|2>$. The opposite effect is generated by a collision occurring in the state $|1>$.

                In conclusion, to take into account the effect of an external perturbation we have to turn Eq. (\ref{bigsigma2}) into

                         \begin{equation}
\label{bigsigmaX}
\Sigma(t) = \left[1 - \frac{S(t-1)}{2}\right] \Sigma(t-1) - \frac{D(t-1)}{2} ,
\end{equation}
where
\begin{equation}
\label{definition1}
D(t) \equiv r_{+}(t) - r_{-}(t)
\end{equation}
and
\begin{equation}
\label{definition2}
S(t) \equiv r_{+}(t) + r_{-}(t).
\end{equation}
The solution of Eq. (\ref{bigsigma}) is
\begin{equation}
\label{stochastic}
\Sigma(t) = -\sum_{t^{\prime}=0}^{t-1} \frac{D(t^{\prime})}{2} Q(t,t^{\prime}),
\end{equation}
where
\begin{equation}
\label{Q1}
Q(t,t^{\prime}) \equiv 1,
\end{equation}
if $t = t^{\prime}+1$,
and
\begin{equation}
\label{Q2}
Q(t,t^{\prime}) \equiv \left[1 - \frac{S(t^{\prime} +1)}{2}\right]...\left[1 - \frac{S(t-1)}{2}\right],
\end{equation}
if $t > t^{\prime}+1$.
Using Eq. (\ref{definition1}), we write Eq. (\ref{stochastic}) as follows
\begin{equation}
\label{stochastic2}
\Sigma(t) = -\sum_{t^{\prime}=0}^{t-1} \frac{r_{+}(t^{\prime})}{2} Q(t,t^{\prime})
+\sum_{t^{\prime}=0}^{t-1} \frac{r_{-}(t^{\prime})}{2} Q(t,t^{\prime}).
\end{equation}

It is important to point out that we prepare the system. This means that $r_{+}(0) = r_{-}(0) = 1$. Thus $\Sigma(1) = 0$. In the continuous time limit this is indistinguishable from $\Sigma(0) = 0$, which is equivalent to assuming that there is no bias on the initial condition. We note that for $t> 1$, the value $\Sigma(t)$ is determined by the events occurring at times $t^{\prime}$ meeting the condition $t-1 \geq t^{\prime} \geq 1$. It is important to notice that $Q(t,t^{\prime}$) depends only the occurrence or non-occurrence of events at times larger than $t^{\prime}$ and smaller than $t$.

The key issue to discuss in the next two sections  is as to how to evaluate
\begin{equation}
\Pi(t) \equiv <\Sigma(t)>.
\end{equation}
Using Eq. (\ref{stochastic2}), we write
\begin{equation}
\label{average}
\Pi(t) = <\Sigma(t)> = - \frac{1}{2}\sum_{t^{\prime}=0}^{t-1} \left [<r_{+}(t^{\prime})Q(t,t^{\prime})> -<r_{-}(t^{\prime})Q(t,t^{\prime})>\right].
\end{equation}
The quantities
$<r_{\pm}(t^{\prime})Q(t,t^{\prime})>$ are evaluated over infinitely many trajectories. All these trajectories begin with the beginning of a quiescent region,  half of them being initially located in the state $|1>$ and half in the state $|2>$. Thus, the quantities $<r_{\pm}(t^{\prime})Q(t,t^{\prime})>$ can be interpreted as being the probabilities that the sequences of events corresponding to the definitions of $r_{\pm}(t^{\prime}) Q(t,t^{\prime})$ really occur. We shall see that these probabilities can be expressed in terms of the quantities $P(t), \Psi(t), \psi(t,t^{\prime})$, all of them being derived from the distribution $\psi(t)$ of Eq. (\ref{theoretical}).


\section{Phenomenological approach}\label{phen}
In this case, we use the prescription of Eq. (\ref{suitablechoice}). Thus Eq. (\ref{average}) yields
\begin{equation}
\label{intuitiveaverage} \Pi(t) =  - \epsilon
\sum_{t^{\prime}=0}^{t-1}
<r_{0}(t^{\prime})Q(t,t^{\prime})>\cos(\omega t^{\prime}).
\end{equation} The time distance between two consecutive events,
namely, two occurrence times where $r_{0}(t)$ is equal to $1$, has
the distribution $\psi(\tau)$, because we are assuming that the
external perturbation does not affect the time distribution of
events but only the rates $r_{+}(t)$ and $r_{-}(t)$ when an event
occurs.

Let us now focus our attention on the function $<Q(t,t^{\prime})>$. We want to prove
\begin{equation}
\label{aneventoccurs}
<Q(t,t^{\prime})> = \Psi(t,t^{\prime}) = \Psi(t-t^{\prime}).
\end{equation}
Note that we rest on the probability that at $t^{\prime}$ an event occurs. If it did not, $r_{0}(t^{\prime}) = 0$, and there would be no contribution to the sum of Eq. (\ref{intuitiveaverage}).  The function
$Q(t,t^{\prime})$, according to the definition of Eqs. (\ref{Q1}) and (\ref{Q2}), is the product of non vanishing factors with the value of $1$, if no event occurs in between $t^{\prime}$ and $t$. In fact, if no event occurs at $t^{\prime}+1$, $S(t^{\prime} + 1) = 0$ and $1-S(t^{\prime} + 1) = 1$. The same argument applies to the following times until time $t-1$. At this time no event occurs either, thereby making $S(t-1) = 0$ and $1-S(t-1)/2 = 1$. Thus, if no event occurs in between $t^{\prime}$ and $t$, as earlier stated, $Q(t,t^{\prime}) = 1$. Let us assume now that an event occurs at $t$. As a consequence $S(t) = 2$. Thus, at the next step we get $1-S(t)/2 = 0$, thereby making $Q(t+1,t^{\prime}) = 0$. Thus $Q(t,t^{\prime})$ is equal to $1$ and it vanishes when we make the change $Q(t,t^{\prime}) \rightarrow Q(t+1,t^{\prime})$.   On the other hand, at $t^{\prime}$ an event occurs. Thus, $<Q(t,t^{\prime})>$, the probability that $Q(t,t^{\prime})$ does not vanish,  is equivalent to the probability that we do not find any event moving from $t^{\prime}$, when an event occurs, up to time $t$. Consequently, not only $<Q(t,t^{\prime})>$ is a survival probability: it is a brand new survival probability, namely evaluated by setting the beginning of the observation at a distance $\Delta t = 1$ from the occurrence of the last event. The measure of this distance becomes virtually zero in the continuous time limit, thereby explaining why we obtain Eq. (\ref{aneventoccurs}).

In the continuous time limit we write
\begin{equation}
\label{firstway} \Pi(t) = -\epsilon \int_{0}^{t} dt^{\prime}
\chi(t,t') \cos(\omega t^{\prime}) ,
\end{equation}
which is the typical linear response function structure \cite{linearesponse,hanggi}. In literature, the function $\chi(t,t^{\prime})$ is called susceptibility. Its explicit expression in the phenomenological case here under discussion is:
\begin{equation}
\label{susceptibility3}
\chi(t,t^{\prime}) = \Psi(t-t^{\prime}) P(t^{\prime}),
\end{equation}
where
\begin{equation}
\label{whore1}
P(t) = \sum_{n=1}^{\infty} \psi_{n}(t) ,
\end{equation}
with $\psi_{n}(t)$ denoting the probability of occurrence, at time $t$, of the last of a sequence of $n$ collisions. Note that the sum is done from $n= 1$ to $\infty$, as a consequence of the fact that $n=0$ corresponds to the preparation event located at $t= 0$. This event does not contribute to the bias, insofar as we assume that at $t=0$ there are as many particles in the state $|1>$ as in the state $|2>$.
Thus, $P(t)$ is the probability of event occurrence at $t$, given the condition that the system is prepared at $t=0$ and the event occurrence is observed at time $t>0$.
The Laplace transform of $P(t)$, $\hat P(u)$, is given by
\begin{equation}
\label{plug}
\hat P(u) = \sum_{n=1}^{\infty} \hat \psi_{n}(u) =
 \sum_{n= 0}^{\infty} [\hat \psi(u)]^{n} - 1 =  \frac{\hat \psi(u)}{1 - \hat \psi(u)},
\end{equation}
where $\hat \psi(u)$ is the Laplace transform of $\psi(\tau) = \psi_{1}(\tau)$.
In the case $\mu < 2$, the Laplace transform of
$\psi(\tau)$ of Eq. (\ref{theoretical}) for $u \rightarrow 0$ has the form
\begin{equation}
\hat \psi(u) = 1 - \Gamma(\mu-1) (uT)^{\mu-1}.
\end{equation}
By plugging this form into Eq. (\ref{plug}) and inverse Laplace transforming the result, we get that for $t\rightarrow \infty$
\begin{equation}
\label{whore}
P(t) \propto \frac{1}{t^{2-\mu}}.
\end{equation}
Thus, the rate of event occurrence decreases as a function of time, an evident aging property that is taken into account by the non-stationary susceptibility of Eq. (\ref{susceptibility3}).

The meaning of the susceptibility $\chi(t,t^{\prime})$ stemming
from the phenomenological condition, is evident. It is
proportional to the probability of event occurrence at $t'$ by the
probability that no further event occurs up to time $t$. Thus the
bias $\Pi(t)$ collects all the collisions occurring at earlier
times and establishing the current $P(t^{\prime}) \cos (\omega
t^{\prime})$. Note that
\begin{equation}
\label{appendixa}
\hat \Pi (u) = - \epsilon \frac{1-\hat \psi(u)}{u} Re\left\{\frac{\hat \psi(u + i\omega)}{[1-\hat \psi(u+i\omega)]}\right\}.
\end{equation}
In Appendix A we illustrate the procedure necessary to inverse Laplace transform this formula in the asymptotic condition $u \rightarrow 0$, so as to evaluate $\Pi(t)$ for $t\rightarrow \infty$ in this case.

\section{Dynamic approach}\label{dynamic}
As noticed in Section \ref{dynamic1}, the dynamic approach derives the system's bias from the fact that, as an effect of perturbation, the events in the state $|1>$ and the events in the state $|2>$ occur at different times.
Let us imagine that two independent trajectories enter the states $|1>$ and $|2>$, respectively, at the same time $t=0$. Thus, at $t=0$ two simultaneous collisions occur in both states. Let us denote by $w_{+}(t^{\prime})$ and $w_{-}(t^{\prime})$ the probabilities that a further collision occurs at time $t^{\prime}$
in the states $|1>$ and $|2>$, respectively. 
Using Eqs. (\ref{puttana}), (\ref{puttanina}) and (\ref{notexact}), we predict that
\begin{equation}
\label{paolowarning} \frac{w_{+}(t^{\prime})}{w_{-}(t^{\prime})} =
\frac{1 + \epsilon \cos (\omega t^{\prime})}{1 - \epsilon \cos
(\omega t^{\prime})}.
\end{equation}
As a consequence of this equality, a collision occurs in the state $|1>$ earlier (later) than
in the state $|2>$, if $\cos (\omega t^{\prime}) > 0$ ($\cos(\omega t^{\prime}) < 0$). This has
the effect of decreasing (increasing) the number of particles in $|1>$ and
increasing (decreasing) the number of particles in $|2>$ by the same amount.

To convert the occurrence of collisions in the two states at different times into a linear response to the external perturbation, let us proceed as follows. Let us set $r_{-}(t) = r_{+}(t) = 1$ and let us write Eq. (\ref{stochastic2}) under the following form
\begin{equation}
\label{stochastic4}
\Sigma(t) = -\sum_{t^{\prime}=0}^{t-1} \frac{r_{-}(t)r_{+}(t^{\prime})}{2} Q(t,t^{\prime})
+\sum_{t^{\prime}=0}^{t-1} \frac{r_{+}(t)r_{-}(t^{\prime})}{2} Q(t,t^{\prime}).
\end{equation}
The first term on the right-hand side of this equation can be interpreted as the occurrence of two events,  a first event occurring in the state $|1>$ at time $t^{\prime}$ and a second event  occurring in the state $|2>$ at a later time $t > t^{\prime}$. In fact, if $r_{+}(t^{\prime}) = 0$ there would be no contribution to this term. This term would also vanish if $r_{-}(t) = 0$. But this would violate the condition that this term coincides with the first term of Eq. (\ref{stochastic2}). With the same argument we prove that the second term on the right-hand side of this equation can be interpreted as corresponding to a first event occurring in the state $|2>$ at time $t^{\prime}$, and to a second event occurring in the state $|1>$ at a later time
$t > t^{\prime}$. We have now to move to the crucial step of making a statistical average. It is evident that in the first term on the right-hand side of Eq. (\ref{stochastic4}) the existence of the terms of the sum over $t^{\prime}$ depends on the probability that $r_{-} (t) = 1$ and $r_{+}(t^{\prime}) = 1$, and that, analogously, in the second term the existence of the terms of the sum over $t^{\prime}$ depends on the probability that $r_{+}(t) = 1$ and $r_{-}(t^{\prime}) = 1$.  Let us make the assumption that no event occurs in between $t^{\prime}$ and $t$, thereby setting $Q(t,t^{\prime}) = 1$.
The average of $\Sigma(t)$ of Eq. (\ref{stochastic4}) yields
\begin{equation}
\label{nocollisioninbetween}
\Pi(t) = - \frac{1}{2} \left[\int_{0}^{t} dt^{\prime} M_{(+ \rightarrow -)}(t^{\prime})- \int_{0}^{t} dt^{\prime} M_{(- \rightarrow +)}(t^{\prime})\right],
\end{equation}
where $M_{(\pm \rightarrow \mp)}(t^{\prime})$ denotes the probability that at time $t^{\prime}$ an event occurs in the state $|1>(|2>)$ and another event occurs at time $t$ in the state $|2>(|1>)$.
Note that in the unperturbed case
\begin{equation}
\label{unperturbed}
M_{(+\rightarrow-)} = M_{(-\rightarrow+)} = \psi(t,t^{\prime}).
\end{equation}
This is so because the occurrence of an event in the state $|1>$ ($|2>$), at time $t^{\prime}$, signals the beginning of observation in the state $|2>$ ($|1>$). In both states an event occur at time $t$ and observation begins at time $t^{\prime}$.
On the other hand, to properly take into account the effect of perturbation, we use Eq. (\ref{paolowarning}). Thus, due to the correlation between event occurrence and quiescent region sign, we obtain
\begin{equation}
\label{perturbed} \frac{M_{(+\rightarrow-)}(t^{\prime})
}{M_{(-\rightarrow+)}(t^{\prime})} =
\frac{w_{+}(t^{\prime})}{w_{-}(t^{\prime})} = \frac{[1+\epsilon
\cos (\omega t^{\prime})]}{[1-\epsilon \cos (\omega t^{\prime})]}.
\end{equation}
To make Eq. (\ref{perturbed}) compatible with Eq. (\ref{unperturbed}) when $\epsilon = 0$, we set
\begin{equation}
\label{faircoin1}
M_{(+\rightarrow -)} = <r_{+}(t) r_{-}(t^{\prime})
Q(t,t^{\prime})> = <r_{0}(t) \Psi(t,t^{\prime})>[1+\epsilon
\cos(\omega t^{\prime})] =\psi(t,t^{\prime})[1+\epsilon
\cos(\omega t^{\prime})]
\end{equation}
and
\begin{equation}
\label{faircoin2}
M_{(-\rightarrow +)} = <r_{-}(t) r_{+}(t^{\prime})
Q(t,t^{\prime})> = <r_{0}(t) \Psi(t,t^{\prime})>[1-\epsilon
\cos(\omega t^{\prime})] =\psi(t,t^{\prime})[1-\epsilon
\cos(\omega t^{\prime})].
\end{equation}
It is important to observe that these two equations correspond to the prescription of Eqs. (\ref{puttana}) and (\ref{puttanina}), involving  Eq. (\ref{notexact}) as well. As pointed out in Section \ref{dynamic}, this prescription can be realized in two different ways, one resting on the unfair coin tossing occurring at the end of the quiescent region, and the other based on the use of a fair coin tossing and of a perturbed waiting distribution. Here we are using the second way of realizing this prescription.

Using Eqs.(\ref{faircoin1}) and (\ref{faircoin2}), we rewrite Eq. (\ref{nocollisioninbetween}) as follows
\begin{equation}
\label{susceptibility1} \Pi(t) = -\int_{0}^{t} dt^{\prime}
\frac{\psi(t,t^{\prime})}{2}[1+ \epsilon \cos(\omega t^{\prime})]
+ \int_{0}^{t} dt^{\prime} \frac{\psi(t,t^{\prime})}{2}[1-
\epsilon \cos(\omega t^{\prime})] = -\epsilon \int_{0}^{t}
dt^{\prime} \chi(t,t^{\prime})\cos(\omega t^{\prime})
,\end{equation} which yields the same form as Eq. (\ref{firstway})
with the susceptibility of Eq. (\ref{susceptibility1}) replaced by
\begin{equation}
\label{susceptibility2}
\chi(t,t^{\prime}) \equiv \psi(t,t^{\prime}).
\end{equation}
This result is based on the assumption that no event occurs in between
$t^{\prime}$ and $t$. On the other hand, this approximation can be expressed in a different way. We made the assumption that the bias at time $t$ depends only the delay between the event on $|1>(|2>)$, occurring at time $t$ and the  event occurring in the other state $|2>(|1>)$ at time $t^{\prime}$. This is equivalent to assuming that at time $t^{\prime}$ there is no bias. If we relax this assumption and consider also the possibility that at time $t^{\prime}$ there is a bias, this is certainly proportional at least to $\epsilon$.  Thus, this correction to the approximation adopted to get Eq. (\ref{susceptibility1}) would be proportional to $\epsilon^{2}$.  In conclusion, the assumption yielding Eq. (\ref{susceptibility2}) is a linear response assumption.

It is interesting to notice that in the Poisson case the prescription of Eq. (\ref{susceptibility1}) coincides with the prescription of Eq. (\ref{firstway}).
Notice that the theoretical result of Eq. (\ref{susceptibility1}) coincides with the prescription of Ref. \cite{prl}.
To make this paper as self-contained as possible, let us review the calculation done in  \cite{prl} to deal with the case  $\mu < 2$, which makes aging a perennial condition of
renewal systems.
It is known \cite{aging1} (see also Ref. \cite{lastgerardo}) that the exact expression for the aged $\psi(t,t^{\prime})$
is
\begin{equation}
\label{aged}
\psi(t,t^{\prime}) = \psi(t) +
\int_{0}^{t^{\prime}} P(\tau) \psi(t-\tau) d\tau,
\end{equation}
where $P(t)$ is the time-dependent rate of event occurrence of Eqs. (\ref{whore1}) and (\ref{whore}), the latter of which makes evident the perennial aging condition of the case $\mu < 2$ here under discussion, and evidently properly considered by Eq. (\ref{susceptibility2}).

It is straightforward to evaluate $\hat \Pi(u)$, the Laplace transform of
 of $\Pi(t)$ of  Eq. (\ref{susceptibility1}). This is so thanks to the time convolved nature of this equation and to Eqs. (\ref{plug})
and (\ref{whore}). We obtain the following expression
\begin{equation}
\label{quote} \hat \Pi(u) = - \epsilon Re (\hat E(u)),
\end{equation}
where $\hat E(u)$ is the Laplace transform of
\begin{equation}
\label{E(t)}
E(t) \equiv \int_{0}^{t} dt^{\prime} \psi(t, t^{\prime}) \exp(- i
\omega t^{\prime}).
\end{equation}
After some algebra, detailed in Appendix B, we find

\begin{equation}
\label{nostepfunction} \hat E(u) = \frac{i}{\omega}
\frac{\left[\hat \psi(u+i\omega) - \hat \psi(u)\right]}{\left[1 -
\hat \psi(u+i\omega)\right ]}.
\end{equation}
In Appendix B we give also details on the procedure adopted to inverse Laplace transform this important equation.

\section{Asymptotic behavior}\label{asym}
We are now in a position to derive the central result of this paper. By means of the calculations illustrated in Appendix A we show that the adoption of the phenomenological theory yields the asymptotic condition
\begin{equation}\label{maurophen}
   \Pi\left(t\right)= \Pi_1\left(t\right)+ \Pi_2\left(t\right) \approx\epsilon
  \left(\frac{T}{T+t}\right)^{\mu-1}-
\frac{\epsilon}{\Gamma(2-\mu)(\omega t)^{
\mu-1}}\sin\frac{\pi\mu}{2}
  +\frac{\epsilon }{\Gamma(\mu-1)}
  \frac{\cos\left(\frac{\pi\mu}{2}+\omega t\right)}{(\omega
  t)^{2-\mu}}.
\end{equation}
In the dynamical case, as shown in Appendix B, we recover
the result of Ref. \cite{prl}:
\begin{equation}\label{maurodyn}
   \Pi\left(t\right) \approx \frac{\epsilon }{\Gamma(\mu-1)}
  \frac{\cos\left(\frac{\pi\mu}{2}+\omega t\right)}{(\omega
  t)^{2-\mu}}.
\end{equation}
Note that in both cases we do not adopt the
 restriction to the condition $\omega T \ll 1$, a constraint used in Ref. \cite{prl}. We have also to point out that we are considering the time asymptotic condition $\omega t \gg1$.

 For $1<\mu<\frac{3}{2}$ the first two terms in the right-hand side of Eq. (\ref{maurophen})  are the
dominant ones, while for $\frac{3}{2}<\mu<2$ the dominant is the
third term.  Thus, when $\mu$ is very close to $1$, the phenomenological theory yields a result with no apparent resonance effects, namely, a significant departure from the prediction of the dynamic approach. On the other hand, we note that tuning the external perturbation to the frequency
\begin{equation}\label{freq}
\omega = \omega_{c} \equiv \left[\frac{\sin\frac{\pi\mu}{2}}{\Gamma(2-\mu)}\right]^{\frac{1}{\mu-1}}\frac{1}{T}
\end{equation}
has the interesting effect,  in the limit
$t\to\infty$, of making the first and second term on the right-hand side of this equation cancel with each other, thereby making Eq. (\ref{maurophen}) become identical to Eq. (\ref{maurodyn}), and the phenomenological identical to the dynamic prediction. At the moment of writing this paper, the physical meaning of this critical frequency is not quite clear.

It is of some interest to discuss the step-like perturbation
\begin{equation}
\label{steplike}
F_{\pm}(t) \equiv \theta(t).
\end{equation}
In both phenomenological and dynamic case the response to a step-like perturbation yields a finite asymptotic value for $\Pi(t)$. In the dynamic case this asymptotic value, though, may become much smaller than in the phenomenological case.
In fact, the phenomenological case yields, see Appendix A,
\begin{equation}
\label{stepfunction1}
\Pi(\infty) = - \epsilon
\end{equation}
and the dynamical case produces, see Appendix B,
\begin{equation}
\label{stepfunction2}
\Pi(\infty) = - \epsilon(\mu-1),
\end{equation}
which vanishes for $\mu \rightarrow 1$.

\section{concluding remarks}\label{final}
This paper affords a detailed account of the dynamical theory proposed in Ref. \cite{prl}. Furthermore, the time asymptotic behavior of the response is evaluated with no restriction to the condition $\omega T \ll 1$, adopted in Ref. \cite{prl}. It is interesting to notice that  with the dynamic approach a form of resonant response is admitted, even if it slowly dies out. With the phenomenological model the non-resonant response may dominate, preventing us from observing any form of resonance.
The response to a step-like perturbation, in the time asymptotic limit yields a finite signal. Note that the time asymptotic value of the response to a step-like perturbation, in the case studied in Ref. \cite{sokolov,barkai} is a vanishing current, in an apparent conflict with the prediction of this paper, where both phenomenological and dynamic approach yield a non-vanishing current, although of different intensity. As pointed out in Section \ref{introduction}, our approach refers to the super-diffusion case, thereby explaining the departure from the results of Ref. \cite{sokolov,barkai}.

In conclusion, we think that this paper, on the one hand, affords the possibility of establishing whether a given system obeys the dynamic or the phenomenological model. Should further investigation, for instance in the field of blinking quantum dots \cite{intermittentfluorescence}, prove that the dynamic model is more plausible than the phenomenological model, the procedure of this paper may be adopted to derive the phenomenological model from dynamics, and to assess under which conditions the phenomenological model can be trusted. It seems that the  approximation adopted  to derive the phenomenological condition from dynamics becomes negligible with $\mu \rightarrow 2$. On the other hand, we think that a challenging problem is to overcome the weak perturbation approach necessary to make theoretical predictions with the dynamic model.

\renewcommand{\theequation}{A-\arabic{equation}}
  \setcounter{equation}{0}  
  \section*{APPENDIX A}  
This appendix is devoted to inverse Laplace transforming Eq. (\ref{appendixa}) that we rewrite here for the reader's convenience:

\begin{equation}\label{mir1}
\hat{\Pi}\left(u \right)=-\epsilon
Re\left[\frac{1-\hat{\psi}(u)}{u}
\frac{\hat{\psi}(u+\imath\omega)}{1-\hat{\psi}(u+\imath\omega)}\right].
\end{equation}
We express the Laplace transform of $\hat\psi(u)$ by means of the exact expression \cite{bolognone,libro}

\begin{equation}\label{mir2}
  \hat{\psi}(u)=\Gamma(1-\mu)\left(Tu
  \right)^{\mu-1}(\mu-1)\left(e^{uT}-E^{uT}_{\mu-1}
  \right),
\end{equation}
where
\begin{equation}
E_{\gamma}^{x} \equiv \sum_{n=0}^{\infty} \frac{x^{n-\gamma}}{\Gamma(n+1-\gamma)}.
\end{equation}

Using Eq. (\ref{mir2}), let us write Eq. (\ref{mir1}) as:

\begin{equation}\label{mir3}
\hat{\Pi}\left(u \right)=-\epsilon
Re\left[\frac{1-\hat{\psi}(u)}{u}\cdot\frac{\hat{\psi}(u+\imath\omega)-1+1}
{1-\hat{\psi}(u+\imath\omega)}\right]=
\epsilon\frac{1-\hat{\psi}(u)}{u}-\epsilon
Re\left[\frac{1}{u}\cdot\frac{1-\hat{\psi}(u)}
{1-\hat{\psi}(u+\imath\omega)}\right].
\end{equation}
Let us set $\hat{\Pi}\left(u \right)=\hat{\Pi}_1\left(u
\right)+\hat{\Pi}_2\left(u \right)$ so that $\hat{\Pi}_1\left(u
\right)=\epsilon\frac{1-\hat{\psi}(u)}{u}$ and $\hat{\Pi}_2\left(u
\right)=-\epsilon Re\left[\frac{1}{u}\cdot\frac{1-\hat{\psi}(u)}
{1-\hat{\psi}(u+\imath\omega)}\right]$.
It is straightforward to inverse Laplace transform  $\hat \Pi_{1}(u)$, which yields $\Pi_1\left(t
\right)=\epsilon\left(\frac{T}{T+t}\right)^{\mu-1}$. The inverse Laplace transform of $\hat{\Pi}_2\left(u \right)$ is more laborious.
First of all, we introduce the notation
$\hat{\Pi}_2\left(u \right)=Re\left[\hat{\Pi}_C\left(u
\right)\right]$ and we rewrite the second term in the right side
of Eq. (\ref{mir3}) as

\begin{equation}\label{mir4}
\hat{\Pi}_C\left(u \right)
=-\frac{\epsilon}{u}(u+\imath\omega)^{1-\mu}
\frac{1-\hat{\psi}(u)} {(u+\imath\omega)^{1-\mu}-
\Gamma(1-\mu)T^{\mu-1}(\mu-1)\left[e^{(u+\imath\omega)T}-E^{(u+\imath\omega)T}_{\mu-1}\right]}
\end{equation}
or

\begin{equation}\label{mir5}
\left\{(u+\imath\omega)^{1-\mu}-
\Gamma(1-\mu)T^{\mu-1}(\mu-1)\left[e^{(u+\imath\omega)T}-E^{(u+\imath\omega)T}_{\mu-1}\right]\right\}
\hat{\Pi}_C\left(u
\right)=-\frac{\epsilon}{u}(u+\imath\omega)^{1-\mu}
\left[1-\hat{\psi}(u)\right].
\end{equation}

 To proceed with our calculations, we have to prove first the important formula \begin{equation}\label{mir6}
 \int\limits^{\infty}_{0}\frac{t^\alpha}{t+T}e^{-u
 t}dt=\frac{\pi}{\sin\pi\alpha}T^\alpha
 \left(E^{uT}_{\alpha}-e^{uT}\right).
\end{equation}
Let us replace the exponential $\exp(-ut)$ on the right-hand side of Eq. (\ref{mir6}) with its Laplace transform, with respect to $u$,
\begin{equation}
\int_{0}^{\infty}   e^{-ut}e^{-up}= \frac{1}{p+t} .
\end{equation}
We are therefore led to the integral \cite{abramowitz}
\begin{equation}
\label{abramowitz} \int_{0}^{\infty} \frac{t^{\alpha}}{t+T}
\frac{1}{t+p} dt = \left(\frac {p^{\alpha}}{p-T} -
\frac{T^{\alpha}}{p-T}\right)\frac{\pi}{\sin(\pi \alpha)}.
\end {equation}
According to the Efros theorem \cite{efros},
the inverse Laplace transform of Eq. (\ref{abramowitz}),
from $p$ to $u$, is the Laplace transform of $t^{\alpha}/(t+T)$.
Using Ref. \cite{libro} we derive from this procedure the right-hand side of Eq. (\ref{mir6}).
This proves the equality of Eq. (\ref{mir6}).

Going back from Eq. (\ref{mir6}) to the time dominion we have:

\begin{equation}\label{mir7}
\frac{1}{\Gamma(\mu-1)}\int\limits^{t}_{0}t'^{\mu-2}e^{-\imath\omega
t' }\Pi_C\left(t-t'
\right)dt'-\frac{1}{\Gamma(\mu-1)}\int\limits^{t}_{0}\frac{t'^{\mu-1}}{t'+T}e^{-\imath\omega
t' }\Pi_C\left(t-t' \right)dt'=g(t),
\end{equation}
where $g(t)$ is the inverse Laplace transform of the right side of
Eq. (\ref{mir5}). After a little algebra:

\begin{equation}\label{mir8}
 \frac{T  }{\Gamma(\mu-1)}
\int\limits^{t}_{0}\frac{t'^{\mu-2}}{t'+T}e^{\imath\omega(t- t')
}\Pi_C\left(t-t' \right)dt'=e^{\imath\omega t }g(t).
\end{equation}

The Laplace transform of the left side of
Eq. (\ref{mir8}) is  $k(u) \hat{\Pi}_C(u-\imath\omega)$,
where $k(u)$ is the Laplace transform of the kernel. As to $u\to 0$,
we write $k(u)=k_0+w(u)$, where

\begin{equation}\label{mir9}
  k_0=\frac{T  }{\Gamma(\mu-1)}\int\limits^{\infty}_{0}
  \frac{t'^{\mu-2}}{t'+T}
dt'=-\frac{\pi}{\sin\pi(\mu)}\frac{T^{\mu-1}}{\Gamma(\mu-1)}.
\end{equation}
We do not write the explicit expression of $w(u)$ since $w(u)\to
0$ for $u\to 0$. The Laplace transform of the right-hand side of
Eq. (\ref{mir8}) is $\hat{g}(u-\imath\omega)$; dropping off the
$\imath\omega$ translation, we study directly the equation for
$\hat{\Pi}_C(u )$ and $\hat{g}(u)$, so that we end up with:

\begin{equation}\label{mir10}
k_0
\hat{\Pi}_C\left(u\right)=-\frac{\epsilon}{u}(u+\imath\omega)^{1-\mu}
\left[1-\hat{\psi}(u)\right].
\end{equation}
Taking the limit $u\to 0$ we arrive at:

\begin{equation}\label{mir11}
\hat{\Pi}_C\left(u\right)= \Gamma(1-\mu)
T^{\mu-1}\frac{(\mu-1)}{k_0}\frac{\epsilon}{u}\left(\frac{u}{u+\imath\omega}\right)^{\mu-1}=
\epsilon\Gamma(1-\mu)
T^{\mu-1}\frac{(\mu-1)}{k_0}u^{\mu-2}\left(\frac{1}{u+\imath\omega}\right)^{\mu-1}.
\end{equation}
It is possible to express Eq. (\ref{mir11}) in terms of a
fractional derivative. We use the definition of fractional
derivative $D^{\alpha}_{\beta}$ of Ref. \cite{libro}, which yields
\begin{equation}\label{mir12}
   {\cal
   L}^{-1}\left[u^{\mu-2}\left(\frac{1}{u+\imath\omega}\right)^{\mu-1}\right]=
   D^{\mu-2}_{t}\left[\frac{e^{-\imath\omega t}t^{\mu-2}}{\Gamma(\mu-1)}\right]
\end{equation}
and, consequently,
\begin{equation}\label{mir13}
   {\cal
   L}^{-1}\left[\hat{\Pi}_C\left(u\right)\right]= \epsilon\Gamma(1-\mu)
T^{\mu-1}\frac{(\mu-1)}{k_0}
  \sum\limits^{\infty}_{n=0} {\mu-2\choose n} D^{\mu-2-n}_{t}\left[\frac{t^{\mu-2}}{\Gamma(\mu-1)}\right]
  D^{n}_{t}\left[ e^{-\imath\omega t}\right].
\end{equation}
Using the expression for $k_0$ of Eq. (\ref{mir9}), after a few calculations we
obtain:
\begin{equation}
\Pi_C\left(t\right)= -\epsilon
  \sum\limits^{\infty}_{n=0} {\mu-2\choose n}
  \frac{(-\imath\omega t)^{n}}{\Gamma(n+1)}
   e^{ -\imath\omega t} =-\epsilon F(2-\mu,1,\imath\omega t)e^{ -\imath\omega t} ,
   \end{equation}
   which can be expressed for $t\to \infty$ as \cite{abramowitz}
 \begin{equation}\label{mir14}
\Pi_C\left(t\right) \approx -\frac{\epsilon}{\Gamma(2-\mu)}(\omega t)^{
1-\mu}e^{\imath\left(\frac{\pi(\mu-1)}{2} \right)}
-\frac{\epsilon}{\Gamma(\mu-1)}(\omega t)^{
\mu-2}e^{-\imath\left(\frac{\pi(\mu-2)}{2}+\omega t\right)}.
\end{equation}
Finally, using the real part of $\Pi_{C}(t)$, we have

\begin{equation}\label{mir15}
   \Pi\left(t\right)= \Pi_1\left(t\right)+ \Pi_2\left(t\right)
    \approx\epsilon  \left(\frac{T}{T+t}\right)^{\mu-1}-
\frac{\epsilon}{\Gamma(2-\mu)(\omega t)^{
\mu-1}}\sin\frac{\pi\mu}{2}
  +\frac{\epsilon }{\Gamma(\mu-1)}
  \frac{\cos\left(\frac{\pi\mu}{2}+\omega t\right)}{(\omega
  t)^{2-\mu}},
\end{equation}
which is the key result of Eq. (\ref{maurophen}).

The response to the step-like perturbation of Eq. (\ref{steplike}) is obtained by the earlier results, by setting $\omega=0$.  In this case

\begin{equation}\label{mir16}
\hat{\Pi}\left(u \right)=-\epsilon \frac{\hat{\psi}(u)}{u}
\end{equation}
and
\begin{equation}\label{mir17}
   \Pi\left(t\right)= - \epsilon\left[
 1- \left(\frac{T}{T+t}\right)^{\mu-1}\right],
\end{equation}
which proves Eq. (\ref{stepfunction1}).

We stress that in this Appendix we  did not set no constraint on $T$ and
$\omega$, as done in the earlier paper of Ref. \cite{prl}, where the condition $\omega T \ll 1$ was adopted. Neither the calculations of Appendix B, made along the same lines of this appendix, are forced to obey the condition $\omega T \ll 1$ of Ref.\cite {prl}.

\renewcommand{\theequation}{B-\arabic{equation}}
  \setcounter{equation}{0}  
  \section*{APPENDIX B}  

In the first subsection of this Appendix we derive Eq. (\ref{nostepfunction}). In the second subsection we show the algebraic details to inverse Laplace transform this equation in the asymptotic limit, and we find the response to the step-like perturbation as well.

\subsection{Analytical Laplace Transform}
Let us make the Laplace transform of Eq. (\ref{E(t)}).
We apply the method of integration by parts, yielding
\begin{equation}
\hat{E}(u)=\int_0^{\infty} e^{-ut} dt \int_0^t\psi(t,t') e^{-i\omega t'} dt'=
\int_0^{\infty} e^{-ut} dt \int_0^t \frac{1}{-i\omega} \frac{d(e^{-i\omega t'})}{dt'} \psi(t,t') dt'.
\end{equation}
This means:
\begin{equation}
\hat{E}(u)=\frac{i}{\omega}
\int_0^{\infty} e^{-ut} dt \left\{\left[\psi(t,t') e^{-i\omega t'}\right]_0^t
- \int_0^t e^{-i\omega t'}\frac{d(\psi(t,t'))}{dt'} dt' \right\}.
\end{equation}
Using the definition of $\psi(t,t')$, given by Eq. (\ref{aged}), we get:
\begin{equation}\label{withomeg}
\hat{E}(u)=\frac{i}{\omega}\int_0^{\infty} e^{-ut} dt \left[ \psi(t,t) e^{-i\omega t}
 -\psi(t) - \int_0^t e^{-i\omega t'} \sum_{n=1}^{\infty}\psi_n(t')\psi(t-t') dt' \right].
\end{equation}
Using again the definition of Eq. (\ref{aged}), we have
\begin{equation}
\label{earlier}
\hat{E}(u)=\frac{i}{\omega}\int_0^{\infty} e^{-ut} dt \left\{ e^{-i\omega t} \left[\psi(t) + \sum_{n=1}^{\infty}\int_0^{t}\psi_n(\tau)\psi(t-\tau) d\tau \right]
-\psi(t) - \int_0^t e^{-i\omega t'} \sum_{n=1}^{\infty}\psi_n(t')\psi(t-t') dt' \right\} .
 \end{equation}
 Taking into account that we are dealing with uncorrelated events ($\hat \psi_{n}(u + i\omega) = (\hat \psi(u + i\omega))^{n} $ ) we write the Laplace transform of Eq. (\ref{earlier}) as follows:
 \begin{equation}
\hat{E}(u)=\frac{i}{\omega} \left\{ \hat{\psi}(u+i\omega) +
\sum_{n=1}^{\infty}\left[\hat{\psi}(u+i\omega)\right]^{n}\hat{\psi}(u+i\omega)
-\hat{\psi}(u) - \sum_{n=1}^{\infty}\left[\hat{\psi}(u+i\omega)\right]^{n}\hat{\psi}(u) \right\}
\nonumber
 \end{equation}
 \begin{equation}
=\frac{i}{\omega} \left\{ \left[\hat{\psi}(u+i\omega)
-\hat{\psi}(u) \right] +
\sum_{n=1}^{\infty}\left[\hat{\psi}(u+i\omega)\right]^{n}\cdot
\left[\hat{\psi}(u+i\omega) -\hat{\psi}(u) \right] \right \}.
\end{equation}
By summation of the geometric series, we obtain the important result of Eq. (\ref{nostepfunction}), written here again for the sake of reader's convenience:
 \begin{equation}
 \label{eq_e}
\hat{E}(u)=\frac{i}{\omega} \cdot \frac{ \hat{\psi}(u+i\omega)
-\hat{\psi}(u)  }{1-\hat{\psi}(u+i\omega)  }.
\end{equation}

As far as the step function case of Eq. (\ref{steplike}) is concerned, we can find the corresponding
Laplace transform with two different methods. The first rests on evaluating the limiting case of
$\omega \rightarrow 0$ of Eq. (\ref{eq_e}). Thus, we obtain:

 \begin{equation}
 \label{eq_edim}
\hat{E}(u)=-\lim_{\omega\to 0}\frac{1}{i\omega} \cdot \frac{
\hat{\psi}(u+i\omega) -\hat{\psi}(u)
 }{ 1-\hat{\psi}(u+i\omega) }= -\frac{
1}{ 1-\hat{\psi}(u )  }  \frac{d}{du}\hat{\psi}(u ) .
\end{equation}
As a second method, let us evaluate the Laplace transform of $ \int_{0}^{t}
 \psi(t, t^{\prime})dt^{\prime}  $, namely, the response generated by the step function, to prove that it yields Eq. (\ref{eq_edim}).
We write
\begin{equation}
\hat{E}(u)=\int_0^{\infty} e^{-ut} dt \int_0^t\psi(t,t')
 dt'=\int_0^{\infty} e^{-ut} dt
 \left\{\left[t'\psi(t,t')\right]_{0}^{t}
 - \int_0^t t'\frac{d(\psi(t,t'))}{dt'} dt'  \right\} .
\end{equation}
By applying a method similar to the one used  to derive Eq. (\ref{withomeg}), we obtain:

\begin{equation}\label{withoutom}
\hat{E}(u)= \int_0^{\infty} e^{-ut} dt  \left[t\psi(t) +t \int_0^t
  \sum_{n=1}^{\infty} \psi_n(\tau)\psi(t-\tau)d\tau
  - \int_0^t
  \sum_{n=1}^{\infty}t'\psi_n(t')\psi(t-t')dt' \right].
 \end{equation}
By making  the Laplace transform of all terms on the right-hand side of Eq. (\ref{withoutom}), we finally arrive at:
\begin{equation}\label{withoutom2}
\hat{E}(u)= -\frac{d}{du}\hat{\psi}(u )
-\frac{d}{du}\left[\hat{\psi}(u )\sum_{n=1}^{\infty}\hat{\psi}(u
)^{n}\right]+
 \hat{\psi}(u ) \frac{d}{du}
  \sum_{n=1}^{\infty}\hat{\psi}(u )^{n }=-\frac{
1}{ 1-\hat{\psi}(u )  }  \frac{d}{du}\hat{\psi}(u ),
 \end{equation}
which is  the expected result. On the basis of this result, we conclude  that the response in time to the step function perturbation can be derived from the general result of the harmonic case for $\omega \rightarrow 0$. Thus, at the end of the next subsection we shall use this argument to find the response in time to the step function perturbation.

\subsection{From the Laplace domain to the ordinary time domain}
At this stage we must make Eq. (\ref{eq_e}) suitable for the inverse Laplace transform procedure.
Therefore we have to find the analytical form of $\hat{\psi}(u)$ for $u\rightarrow 0$.
Let us use again the same departure point as in Appendix A, namely Eq. (\ref{mir2}),
which we rewrite here for the sake of reader's convenience
\begin{equation}\label{psilp}
  \hat{\psi}(u)=\Gamma(1-\mu)\left(Tu
  \right)^{\mu-1}(\mu-1)\left(e^{uT}-E^{uT}_{\mu-1}
  \right).
\end{equation}
With the help of this important property, we rewrite Eq. (\ref{eq_e}) under the following form:
\begin{equation}\label{eq_e1}
\hat{E}\left(u\right)=\frac{\imath}{\omega}(u+\imath\omega)^{1-\mu}
\frac{\hat{\psi}(u+\imath\omega)-\hat{\psi}(u)}
{(u+\imath\omega)^{1-\mu}-
\Gamma(1-\mu)T^{\mu-1}(\mu-1)\left[e^{(u+\imath\omega)T}-E^{(u+\imath\omega)T}_{\mu-1}\right]}.
\end{equation}
From this equality we get:
\begin{equation}\label{eq_e2}
\left\{(u+\imath\omega)^{1-\mu}-
\Gamma(1-\mu)T^{\mu-1}(\mu-1)\left[e^{(u+\imath\omega)T}-E^{(u+\imath\omega)T}_{\mu-1}\right]\right\}
\hat{E}\left(u\right)=\frac{\imath}{\omega}(u+\imath\omega)^{1-\mu}
\left[\hat{\psi}(u+\imath\omega)-\hat{\psi}(u)\right].
\end{equation}
For the sake of clarity we rewrite Eq. (\ref{mir6}):

\begin{equation}\label{lp1}
 \int\limits^{\infty}_{0}\frac{t^\alpha}{t+T}e^{-u
 t}dt=\frac{\pi}{\sin\pi\alpha}T^\alpha
 \left(E^{uT}_{\alpha}-e^{uT}\right).
\end{equation}
Following the same procedure  of Eq. (\ref{mir7}), going back to the time dominion, we have:
\begin{equation}\label{e_time}
\frac{1}{\Gamma(\mu-1)}\int\limits^{t}_{0}t'^{\mu-2}e^{-\imath\omega
t' }E\left(t-t'
\right)dt'-\frac{1}{\Gamma(\mu-1)}\int\limits^{t}_{0}\frac{t'^{\mu-1}}{t'+T}e^{-\imath\omega
t' }E\left(t-t' \right)dt'=g(t),
\end{equation}
where $g(t)$ is the inverse Laplace transform of the right side of
Eq. (\ref{eq_e2}). After a little algebra, we arrive at:

\begin{equation}\label{e_time2}
 \frac{T  }{\Gamma(\mu-1)}
\int\limits^{t}_{0}\frac{t'^{\mu-2}}{t'+T}e^{\imath\omega(t- t')
}E\left(t-t' \right)dt'=e^{\imath\omega t }g(t).
\end{equation}
The Laplace transform of the left-hand side of Eq. (\ref{e_time2})
yields $k(u) \hat{E}(u-\imath\omega)$, where $k(u)$ is the Laplace
transform of $t^{\mu-2}/(t+T)$. In the limiting case $u\to 0$ we
write $k(u)=k_0+w(u)$, with
\begin{equation}\label{k}
  k_0=\frac{T  }{\Gamma(\mu-1)}\int\limits^{\infty}_{0}
  \frac{t'^{\mu-2}}{t'+T}
dt'=-\frac{\pi}{\sin\pi(\mu)}\frac{T^{\mu-1}}{\Gamma(\mu-1)}.
\end{equation}
We do not write the explicit expression of $w(u)$, which tends to
vanish for $u\to 0$. The Laplace transform of the right-hand side
of Eq. (\ref{e_time2}) is $\hat{g}(u-\imath\omega)$. We do not
make any additional approximation, and we study directly the
equation for $\hat{E}(u )$ and $\hat{g}(u)$. After a few
calculations we get:

\begin{equation}\label{emark}
k_0
\hat{E}\left(u\right)=\frac{\imath}{\omega}(u+\imath\omega)^{1-\mu}
\left[\hat{\psi}(u+\imath\omega)-\hat{\psi}(u)\right].
\end{equation}
Using Eq. (\ref{psilp}) we finally obtain:

\begin{equation}\label{prl30}
\hat{E}\left(u\right)=\Gamma(1-\mu)
T^{\mu-1}\frac{(\mu-1)}{k_0}\frac{\imath}{\omega}
\left[e^{(u+\imath\omega)T}-E^{(u+\imath\omega)T}_{\mu-1}-
\left(\frac{u}{u+\imath\omega}\right)^{\mu-1}
\left(e^{uT}-E^{uT}_{\mu-1}\right)\right].
\end{equation}

We stress that so far only the function $k(u)$ is approximated
while the right side of Eq. (\ref{emark}) is exact and we do not
set any constraint on $T$ and $\omega$, as done in Ref. \cite{prl}.
From Eq. (\ref{prl30}),
taking the limit for $u\to 0$ and $\omega\to 0$  we obtain:
\begin{equation}
 \label{prl31}
\hat \Pi(u) =  \epsilon Re \left\{\frac{i}{\omega} \left[1 -
\left(\frac{u}{u+i\omega}\right)^{\mu-1}\right] \right\}.
\end{equation}
 This equation was found in Ref. \cite{prl}, under the assumption $\omega T \ll 1$.
 Let us invert Eq. (\ref{prl30}) term by term.
Using Eq. (\ref{lp1}), we obtain:

\begin{equation}\label{Iterm}
   {\cal
   L}^{-1}\left[e^{(u+\imath\omega)T}-E^{(u+\imath\omega)T}_{\mu-1}\right]=
  - \frac{\sin \pi (\mu-1)}{\pi T^{\mu-1}}\frac{t^{\mu-1}}{t+T}e^{-\imath\omega t}
\end{equation}
and
\begin{equation}\label{IIterm}
   {\cal
   L}^{-1}\left[\left(\frac{u}{u+\imath\omega}\right)^{\mu-1}
\left(e^{uT}-E^{uT}_{\mu-1}\right)\right]=
  \frac{1 }{\Gamma(\mu-1)\Gamma(1-\mu)}\int\limits^{t}_{0}
  \frac{t'^{\mu-2}}{(t-t'+T)^{\mu}}e^{-\imath\omega t'}dt'.
\end{equation}
Let us develop  the exponential factor of the right-hand term of Eq. (\ref{IIterm})
into a time series, and let us evaluate the corresponding integrals.  This yields:

\begin{equation}\label{IIbis}
   {\cal
   L}^{-1}\left[\left(\frac{u}{u+\imath\omega}\right)^{\mu-1}
\left(e^{uT}-E^{uT}_{\mu-1}\right)\right]=
  \frac{(\mu-1) t^{\mu-1}\sin\pi\mu}{\pi (t+T)^{\mu}}
  \sum\limits^{\infty}_{n=0}
  \frac{(-\imath\omega t)^{n}}{(n+\mu-1)n!}F\left(\mu,n+\mu-1,n+\mu,\frac{t}{t+T}\right),
\end{equation}
where $F(\alpha,\beta,\gamma,z)$ is the hypergeometric function.
Using the results of Eqs. (\ref{Iterm}) and (\ref{IIbis}), after a
little algebra, we get for $E(t)$ the following expression:

\begin{equation}\label{pretot}
   E(t)=
  \frac{\imath  }{\omega }
  \left[-\frac{(\mu-1) t^{\mu-1}\sin\pi\mu}{\pi (t+T)^{\mu}}
  \sum\limits^{\infty}_{n=1}
  \frac{(-\imath\omega t)^{n}}{(n+\mu-1)n!}F\left(\mu,n+\mu-1,n+\mu,\frac{t}{t+T}\right) \right]
   -\frac{\sin\pi\mu}{\pi}\frac{1-e^{-\imath\omega t}}{ \imath\omega T}
 \left(\frac{t}{T}\right)^{\mu-1}\frac{T}{t+T}.
\end{equation}
In the time asymptotic limit $t\to\infty$, it is possible to evaluate the sum appearing in Eq. (\ref{pretot}). This yields:

\begin{equation}\label{etot}
   E(t)=\left(\frac{t}{t+T}\right)^{\mu}\left[
    e^{-\imath\omega t} (\mu-1)F(2-\mu,2,\imath\omega t)+
    \frac{\sin\pi\mu}{\pi}
 \frac{1-e^{-\imath\omega t}}{ \imath\omega T}\left(\frac{t}{T}\right)^{\mu-2}\right]
 -\frac{\sin\pi\mu}{\pi}\frac{1-e^{-\imath\omega t}}{ \imath\omega T}
 \left(\frac{t}{T}\right)^{\mu-2}\frac{t}{t+T},
\end{equation}
where $F(\alpha,\gamma,z)$ is the confluent hypergeometric. The
first term between the square brackets, multiplied by $-\epsilon$, coincides with Eq. (34) of Ref. \cite{prl}. We know \cite{prl} that for $t\to\infty$ $F(2-\mu,2,\imath\omega t)\sim\frac{1}{t^{2-\mu}}$ yielding Eq. (36) of Ref. \cite{prl}. We now prove that this earlier result can be recovered without setting the condition $\omega T \ll 1$ adopted in Ref. \cite{prl}.
In the time asymptotic limit, Eq. (\ref{etot}) yields:
\begin{equation}
\label{etot2}
   E(t)\approx(\mu-1)F(2-\mu,2,\imath\omega t) e^{-\imath\omega
   t}-\frac{\sin\pi\mu}{\pi}\frac{1-e^{-\imath\omega t}}{ \imath\omega T}
    \left(\frac{t}{T}\right)^{\mu-3}.
\end{equation}
For $\omega\to 0$ the second term of
the right side of Eq. (\ref{etot2}) can give a non-negiglible contribution in a
certain region of time, more precisely when $t\to\infty$ but still with
$\omega t\ll 1$. In fact,  in this case the exponential function in the second
term of the right-hand side of Eq. (\ref{etot2}) produces a factor of the order $t/T$.
Moving to  $\Pi(t)$ we obtain:

\begin{equation}
\label{prefin}
 \Pi(t)\approx
 -\epsilon
\textrm{Re}\left[(\mu-1)F(2-\mu,2,\imath\omega t) e^{-\imath\omega
   t}\right]
 +\epsilon\frac{\sin\pi\mu}{\pi}\frac{\sin\omega t}{\omega T}
    \left(\frac{t}{T}\right)^{\mu-3},
\end{equation}
with a dependence on $T$.
For  $\omega t\gg 1$, taking into account that $\mu <2$ and the asymptotic properties of the confluent hypergeometric function \cite{abramowitz} as well, we get:

\begin{equation}
\label{fin}
 \Pi(t)\approx\epsilon
\frac{\cos\left(\frac{\pi}{2}\mu+ \omega t
 \right)}{\Gamma(\mu-1)(\omega t)^{2-\mu}},
\end{equation}
which coincides with Eq. (36) of Ref. \cite{prl} without involving the condition $\omega T \ll 1$.
Note that Eq. (\ref{fin}) is the important result of  Eq. (\ref{maurodyn}),
whose derivation is one of the aims of this appendix.

As to the step-like perturbation in this case, we proceed as follows.  Let
us examine the case (see Eq. (\ref{eq_edim})):
\begin{equation}\label{omeg0}
\hat{E}(u)=-\frac{ 1}{ 1-\hat{\psi}(u )  }
\frac{d}{du}\hat{\psi}(u ).
\end{equation}
Using the Taylor series for $\hat{\psi}(u )$, we have (for
simplicity we set $T=1$)

\begin{equation}\label{psilp2}
 \hat{\psi}(u)=\Gamma(1-\mu)
 u^{\mu-1}(\mu-1)\left[1+u-\frac{u^{1-\mu}}{\Gamma(2-\mu)}-
 \frac{u^{2-\mu}}{\Gamma(3-\mu)}
 \right].
\end{equation}
Plugging into Eq.(\ref{omeg0})and keeping the lowest orders we
obtain:

\begin{equation}\label{omeg0_2}
\hat{E}(u)\approx \frac{\mu-1}{u}+\frac{c}{u^{\mu-1}},
\end{equation}
corresponding to

\begin{equation}\label{omeg0_t}
E(t)\approx  \mu-1 +\frac{k}{t^{2-\mu}}.
\end{equation}
Let us make a check to confirm that $E(t)\to  \mu-1$ for $t\to\infty$.
Let us go back from  Eq.(\ref{omeg0}) to time dominion.
\begin{equation}\label{time}
 E(t)-\int\limits_{0}^{t}E(t-\tau)\psi(\tau )d\tau=
 t\psi(t).
\end{equation}
So far we did not make any approximation. Let us check if $E(t)\approx \mu-1$
is a solution of Eq. (\ref{time}). We write
\begin{equation}
\label{check}
 \left(\mu-1\right)\left[1-\int\limits_{0}^{t}
 \frac{\mu-1}{(\tau+1)^{\mu}}d\tau\right]= \left(\mu-1\right)\frac{t}{(t+1)^{\mu}},
\end{equation} which in the limit of $t\to \infty$ make Eq.(\ref{check}) become an identity, thereby proving Eq. (\ref{stepfunction2}).

\emph{Acknowledgments}
 We are indebted to professor Steve Shore for the critical reading of the manuscript. PG thankfully acknowledges the Welch foundation for financial support
 through grant \# 70525.

\end{document}